\PassOptionsToPackage{unicode}{hyperref}
\PassOptionsToPackage{naturalnames}{hyperref}
\documentclass[preprint]{elsarticle}

\newcommand{\countOfFigures}{9}
\newcommand{\countOfTables}{10}
\newcommand{\countOfDatasets}{12}
\newcommand{\countOfArchitectures}{22}
\newcommand{\countOfReviewedPapers}{38}

\newcommand{\reviewYears}{2015-2020}
\newcommand{\reviewYearsAnd}{2015 and 2020}

\usepackage{graphicx}
\usepackage{adjustbox}
\usepackage{booktabs}
\usepackage{multirow}
\usepackage[section]{placeins}

\usepackage{todonotes}

\usepackage{hyperref}
\hypersetup{colorlinks = true,linkcolor = blue,anchorcolor =red,citecolor = blue,filecolor = red,
            pdfauthor="Atilla Ozgur"}

\journal{Image and Vision Computing}  %

\begin{document}    %

\begin{frontmatter}

\title{A systematic review of transfer learning based approaches for diabetic retinopathy detection}



\author[baskent-bioe]{Burcu Oltu}

\author[baskent-bioe]{Büşra Kübra Karaca}

\author[baskent-eee]{Hamit Erdem}

\author[jacobs]{Atilla Özgür}
\cortext[jacobs]{Corresponding author}
\ead{a.oezguer@jacobs-university.de}

\address[baskent-bioe]{Baskent University, Department of Biomedical Engineering, Baglica Campus, 06790, Ankara, Turkey}
\address[baskent-eee]{Baskent University, Department of Electrical and Electronics Engineering , Baglica Campus, 06790, Ankara, Turkey}
\address[jacobs]{Jacobs University, Department of Mathematics and Logistics, Campus Ring 1, 28759 Bremen, Germany}

\begin{abstract}
Cases of diabetes and related diabetic retinopathy (DR) have been increasing at an alarming rate in modern times. 
Early detection of DR is an important problem since it may cause permanent blindness in the late stages. 
In the last two decades, many different approaches have been applied in DR detection. 
Reviewing academic literature shows that deep neural networks (DNNs) have become the most preferred approach for DR detection.
Among these DNN approaches, Convolutional Neural Network (CNN) models are the most used ones in the field of medical image classification.
Designing a new CNN architecture is a tedious and time-consuming approach.
Additionally, training an enormous number of parameters is also a difficult task.  
Due to this reason, instead of training CNNs from scratch, using pre-trained models has been suggested in recent years as transfer learning approach. 
Accordingly, the present study as a review focuses on DNN and Transfer Learning based applications of DR detection considering \countOfReviewedPapers{} publications between \reviewYearsAnd{}.
The published papers are summarized using \countOfFigures{} figures and \countOfTables{} tables, giving information about \countOfArchitectures{} pre-trained CNN models, \countOfDatasets{} DR data sets and standard performance metrics.
\end{abstract}

\begin{keyword}
deep learning \sep convolutional neural networks \sep diabetic retinopathy \sep transfer learning \sep pre-trained
\end{keyword}

\end{frontmatter}


\section{Introduction}



Diabetes is one of the most common diseases in the modern world.
According to a report by World Health Organization \cite{WHO201610}, more than 400 million people have diabetes in the world.
By 2024, this number is expected to increase to 552 million  \cite{Guariguata2014Global}.
Diabetes is an important cause for death, blindness, and amputation, according to the same report \cite{WHO201610}.
Diabetic Retinopathy (DR), as one of the common eye diseases caused by diabetes, is a leading cause of blindness \cite{WHO201610, Bourne2013Causes}.

From the point of view of health care, it is more effective when diabetes is detected at an early stage \cite{Safi2018Early}. 
Applications of Artificial intelligence (AI), machine learning (ML), and Deep Neural Networks (DNNs)  in health care have become more common in the last decades.
Many ML based approaches have been applied for detecting, analyzing, and classifying DR images \cite{Safi2018Early, Ishtiaq2019Diabetic, Stolte2020survey}. 
DNNs are a subset of ML methods, and the convolutional neural networks (CNNs) are a subset of DNN methods \cite{LeCun2015Deep}.
CNNs, which simulate the human visual system and are capable of extracting features automatically, have been successfully used for image-based classification and pattern recognition problems, such as lip reading \cite{FernandezLopez2018Survey}, video anomaly detection \cite{Nayak2021comprehensive}, fruit classification \cite{Hameed2018comprehensive}.
DNNs and CNNS have also been applied to many different problems in health care \cite{Lin2016Neural, Shamshirband2021review, Asiri2019Deep,Trokielewicz2020Post}.
CNNs as a supervised learning architecture is the most applied method in image-based classification \cite{Alyoubi2020Diabetic, Badar2020Application}.
CNNs were also used to tackle the DR detection task during the last decade \cite{Asiri2019Deep, Nielsen2019Deep, Sarki2020Automatic}.

CNNs have superior performance, but in order to train them, a lot of time and huge datasets are required \cite{Kandel2020Transfer}. 
Since only a limited number of images are available in the medical image classification tasks, training DNN algorithms are a challenge. 
In order to overcome this shortcoming, applying transfer learning (TL) methods have been proposed.
Transfer Learning can be defined as learning a new task through the transfer of knowledge from an already learned related task.
Recent studies have shown that Transfer Learning approaches do not need large datasets. 
Additionally, the required training times are also reduced since models are already somewhat pre-trained. \cite{Alyoubi2020Diabetic}.

Considering the popularity of the DR and DNNs, many review studies have been published \cite{Kandel2020Transfer,Stolte2020survey,Asiri2019Deep,Alyoubi2020Diabetic,Nielsen2019Deep,Ishtiaq2019Diabetic,Badar2020Application,Sarki2020Automatic,Chu2020Essentials}.
Details of these studies could be found in section~\ref{section-related-works}.
The only existing work on the intersection of DR, DNN, and TL is the study of Kandel and Castelli \cite{Kandel2020Transfer}.
This fast-moving field requires more review studies for DR, DNNs, and TL approaches; therefore, the current review study is prepared.
Considering previously published review articles in the same field, the present study differs from the aforementioned studies due to the following aspects:

\begin{enumerate}
	\item Considering three main keywords, DR, DNN, and TL, this study reviews \countOfReviewedPapers{} articles which are published in the period between \reviewYears{}. 

    \item These details of reviewed studies are summarized in \countOfFigures{} figures and \countOfTables{} tables.

	\item These \countOfReviewedPapers{}  articles are categorized into \countOfArchitectures{} CNN based architectures in Table~\ref{table-architectures}.

    \item Used datasets (\countOfDatasets{} of them) have been analyzed considering their size and number of classes, in reviewed studies, Table~\ref{table-dataset-study} and Table~\ref{table-dataset-size-link}.

    \item Most of the reviewed articles compared their proposed performances using similar performance metrics. These metrics have been shown in Table~\ref{table-performancemetrics-study}.

    \item Additionally, the achieved accuracy and AUC values due to these \countOfDatasets{} datasets have been presented in Table~\ref{table-dataset-acc-auc}.

    \item Some of the reviewed studies have reported their used DNN optimization algorithms. These optimization algorithms have been given and discussed in Table~\ref{table-optimizer-study}.

    \item Pre-processing and data augmentation techniques applied in the reviewed articles have been evaluated and summarized in Table~\ref{table-preprocessing} and Table~\ref{table-augmentation}.

\end{enumerate}

The findings of this systematic review  would enable the researchers to find more easily which databases, algorithms and performance metrics used in DR studies.


\section{Related Works}
\label{section-related-works}

In 2019, Asiri et al.\cite{Asiri2019Deep} reviewed studies that applied various DNN methods employed for DR diagnosis.
In their study, detailed information about DR is given, and 18 DR datasets that can be employed for 4 different DR detection tasks are introduced. 
Nielsen et al. \cite{Nielsen2019Deep} reviewed 11 studies to evaluate the diagnostic accuracy of DNN to classify DR images. 
They summarized the characteristics and results of the reviewed studies,  datasets used, and participants included in the studies using 2 tables. 
Ishtiaq et al. \cite{Ishtiaq2019Diabetic} published a detailed review on the application of AI techniques for the detection of DR, reviewing 74 papers from eight academic databases. 
Reviewed studies are discussed in various views such as datasets, image pre-processing methods, ML approaches, DNN based approaches, and evaluation of performance. 

In 2020, Alyoubi et al.\cite{Alyoubi2020Diabetic} published a review that analyzed the recent state-of-the-art DNN techniques for DR image detection and classification. 
They reviewed 33 studies due to used image pre-processing techniques, used screening systems (binary, multi-level, lesion-based, vessel-based), and 13 publicly available datasets. 
Stolte et al.\cite{Stolte2020survey} performed a comprehensive survey on medical image analysis in DR that provides a description of the currently used technology for DR detection in extensive detail. 
In their study, a detailed introduction about DR, current technologies, and available resources is given beside a discussion of frameworks used for DR detection and classification. 

Also, in 2020, Badar et al. \cite{Badar2020Application} presented a review study considering the application of DNN for the detection of retinal image impairments like DR. 
The published studies were analyzed due to publicly available datasets contains fundus and optical coherence tomography retinal images and DNN architectures. 
Lastly, they showed that DNN capable of replacing traditional classification methods.
Another review study that combines DNN and DR published by Chu et al. \cite{Chu2020Essentials} researched 15 high-quality papers. 
These papers are analyzed according to the selected dataset and performance criteria. 
Moreover, different from other review studies, this article focuses on the potential algorithm limitations of each study.
Sarki et al. \cite{Sarki2020Automatic} published a review article for analyzing detection of Diabetic Eye Disease (DED) using DNN. 
DED includes DR and three various eye diseases, and 65 articles are reviewed from 10 academic databases in their study. 
Datasets, image processing methods, and detailed classification approaches (Tranfer learning, DNN, and combined DNN and ML classification) used for each eye disease are presented, respectively. 
Consequently, they show that DNN provides valuable results for DED detection.

Among the published review studies, the only study that focuses on DR classification by transfer learning is published by Kandel and Castelli \cite{Kandel2020Transfer} in 2020. 
In their study, 18 papers were analyzed in accordance with architectures used, datasets used, optimizers used, and fine-tuning technique used. 
Ultimately, they showed that transfer learning can provide a substantial contribution to DR detection and classification.

\section{Overview of Methodology}
\label{section-proposed-method}
\label{proposed-method}

\begin{figure}[ht!]
  \includegraphics[scale=0.5]{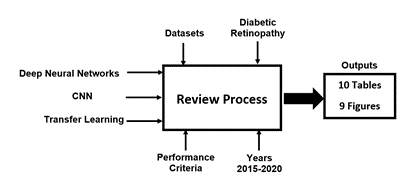}
  \centering
  \caption{Block diagram of review process.}
  \label{figure-blockdiagram}
\end{figure}

\begin{figure}[ht!]
  \includegraphics[scale=0.32]{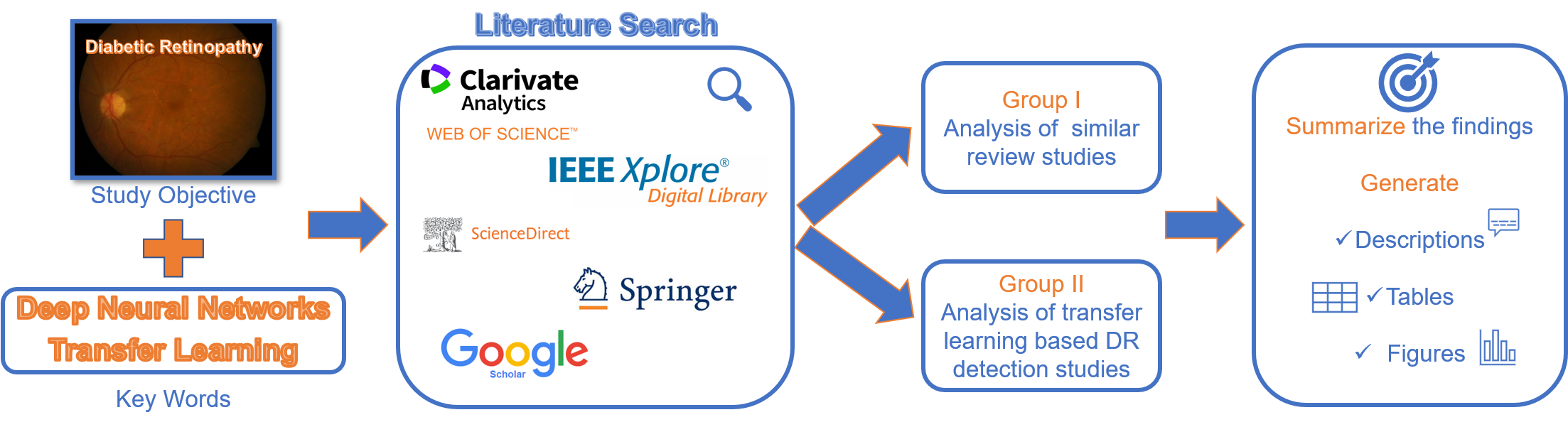}
  \centering
  \caption{Steps of research.}
  \label{figure-stepsofresearch}
\end{figure}

The review process that was adapted for this study is shown in Figure~\ref{figure-blockdiagram}. 
Initially, the review keywords were identified as "Diabetic Retinopathy, Deep Learning, and Transfer Learning".
Then search keywords were used to find the most relevant studies from various academic publications between \reviewYearsAnd{}.
As given in Figure~\ref{figure-stepsofresearch}, the main filter was applied to select the related articles in two main groups: similar review studies and CNN-based DR detection studies.  
In the next step, as given in the process block diagram, due to 3 inputs, \countOfTables{} table and \countOfFigures{} figures as outputs have been generated. 
The details and discussion of the provided tables and figure have been presented under related titles.

\section{Background Materials}
\label{section-background-materials}

\subsection{Diabetic Retinopathy}
\label{section-diabetic-retinopathy}

Diabetic retinopathy (DR) is a complication of diabetes that causes vision problems and blindness. 
DR related blindness is preventable when it is early and correctly detected. 
According to World Health Organization \cite{WHO201610}, the global incidence of diabetes is projected to reach 552 million by 2024.

DR is classified into 5 stages according to the severity level.
Stage 0 refers to non-apparent retinopathy, stage 1 is mild None-Proliferative DR (NPDR), stage 2 is Moderate NPDR, stage 3 is Severe NPDR, and stage 5 is Proliferative DR \cite{Wang2018Diabetic}. 
NPDR is the stage where the proliferative process has not yet begun, and patients are asymptomatic. 
At the NPDR stage, the visual acuity of the patients continues.
Therefore, determining the NPDR stage of the patient is essential in predicting the risk of progression to proliferative retinopathy. 
Early and accurate detection of DR is necessary to prevent the progression of the disease and reduce the risk of vision loss.
Early detection plays a vital role in optimal treatment \cite{Wan2018Deep, Li2019Automatic}. 
The traditional diagnosis of DR is made by manual examination of retinal scanning \cite{Bodapati2020Blended, Li2019Automatic}. 
This diagnostic method is a time-consuming task and the result obtained depends on the experience of the clinician \cite{Wan2018Deep, Li2019Automatic}.

\subsection{Deep Neural Networks and Transfer Learning}
\label{section-DNN-transfer-learning}

Terms of deep learning, deep neural networks, and deep neural nets are used interchangeably in the literature \cite{LeCun2015Deep}. 
DNNs are an advanced version of neural networks that belong to the family of ML and AI methods. 
Unlike traditional neural networks, DNNs have many hidden layers for low-level feature extraction. \cite{Kandel2020Transfer}. 
Over the years, in the context of DNNs, various architectures have been introduced. 
Among these architectures, CNNs, which are feed-forward, multi-layered neural networks composed of feature extractor and classifier parts, have high achievements, especially on medical image classification \cite{Chouan2020Novel}. 
Basic CNN architecture consists of four main parts: convolutional layers, pooling layers, fully connected (FC) layers, and a decision layer (output) \cite{Abideen2020Uncertainty}.

\begin{figure}[ht!]
  \includegraphics[scale=0.5]{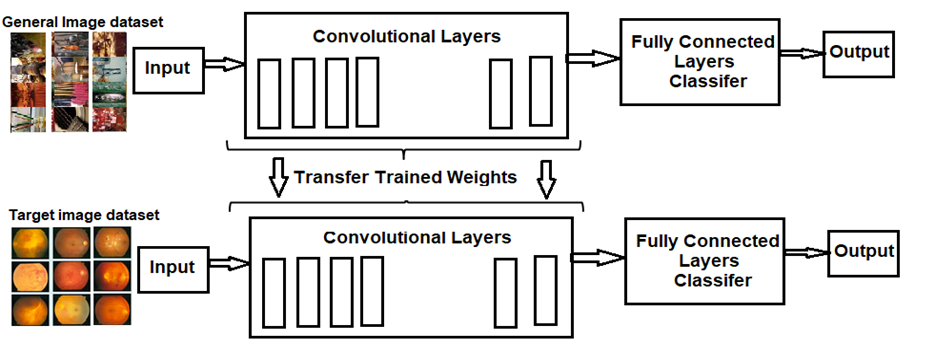}
  \centering
  \caption{Transfer learning approach.}
  \label{figure-transfer-learning-block-diagram}
\end{figure}

The most crucial step of CNN applications is the training process. 
In order to avoid under-fitting and over fitting, and expand generalization, CNNs need to be trained appropriately. 
CNN can be trained with 2 different approaches: training from scratch and transfer learning. 
The transfer learning approach has three strategies. 
In the first three strategies, original FC layers are removed. 
Accordingly, the first strategy is to use the CNN pre-trained layers as feature extractor without any change and add a new classifier layer instead of the original FC layers. 
The second strategy is to fine-tune the entire network and add a suitable classifier layer for the task. 
The third strategy is to fine-tune only the selected amount of top layers where the bottoms layers are frozen and then to add a new, suitable classifier layer. 
An additional training strategy is sometimes mentioned as transfer learning approach.
This last strategy is to select the state-of-the-art pre-trained structure and to train the architecture without any addition or removal (Figure~\ref{figure-transfer-learning-block-diagram}).
This strategy is also same as training from scratch.
Except for the last strategy, the transfer learning approaches do not require a large amount of training data.
Since most medical image datasets are not very large, applying transfer learning for image-based medical diagnosis is beneficial.

Many of the popular pre-trained CNN models are trained using ImageNet \cite{Kieu2020Suvey}. 
ImageNet is a large-scale image classification dataset that contains 14.197.122 annotated images with 1000 object classes. 
Annually, an object classification and detection competition called The ImageNet Large Scale Visual Recognition Challenge (ILSVRC) uses a subset of ImageNet is run \cite{Russakovsky2015Imagenet}. 
Over the years, distinct approaches and architectures were presented in this competition. 
However, CNN architectures have been continuous winners of this competition since 2012. Even the dominance of CNN structures for image classification has been so well recognized that almost all of the structures participating in the competition in recent years are CNN structures. 
Although the basic components (e.g., convolutional and pooling layers) of the models are almost identical in newly presented architectures, some topological differences are presented in modern architectures \cite{Alom2019State}. 
Accordingly, since the introduction of AlexNet many highly thriving architectures, such as VGG, ResNet, DenseNet, GoogleNet, were presented for image classification. \cite{Kandel2020Transfer}. 
Therefore, many of these well-known pre-trained CNN architectures have also been used for DR detection.

In the reviewed studies, various transfer learning approaches were preferred as can be seen in Table~\ref{table-transfer-learning-approaches}. 
In some of the studies, the preferred approach was given in detail, however in most of the studies the preferred strategy is just stated as Transfer Learning and details were not given. 
Note that if the study mentioned the used strategy as transfer learning or training a pre-trained network it is evaluated as fine-tuning the entire network in Table~\ref{table-transfer-learning-approaches}.

\begin{table}[!htb]
\centering
\caption{Transfer learning strategies used in the review studies.}
\label{table-transfer-learning-approaches}
\resizebox{\textwidth}{!}{%

\begin{tabular}{p{0.5\textwidth}p{0.5\textwidth}}

\toprule

TL strategy & Study \\
\midrule

Using pre-trained network without fine-tuning & 
\cite{Qomariah2019Classification},
\cite{Li2019Automatic} \\

Fine-tuning entire pre-trained network & 
\cite{Vo2016New},
\cite{Choi2017Multi},
\cite{Lam2018Retinal},
\cite{Wang2020Multi},
\cite{Li2017Convolutional},
\cite{Zhou2018Deep},
\cite{Lian2018Deep},
\cite{Gao2018Diagnosis},
\cite{Lam2018Automated},
\cite{Wan2018Deep},
\cite{Wang2018Diabetic},
\cite{Xu2018Improved},
\cite{Hathwar2019Automated},
\cite{Kassani2019Diabetic},
\cite{Wijesinghe2019Transfer},
\cite{Li2019Diagnostic},
\cite{Ahmad2019Deep},
\cite{Math2019Identification},
\cite{Yip2019Enhanced},
\cite{Dekhil2019Deep},
\cite{Islam2019DeepDR},
\cite{Hagos2019Transfer},
\cite{Voets2019Reproduction},
\cite{Zhang2019Automated},
\cite{Bodapati2020Blended},
\cite{Mateen2020Exudate},
\cite{Roshan2020Fine},
\cite{Yu2020Performance},
\cite{Hacisoftaoglu2020Deep}\\

Fine-tuning a part of the pre-trained network & 
\cite{Mohammadian2017Comparative},
\cite{Wang2020Multi},
\cite{Li2017Convolutional} \\

Training a state-of-art architecture from scratch & 
\cite{Takahashi2017Applying},
\cite{Wang2020Multi},
\cite{Lian2018Deep},
\cite{Xu2018Improved} \\

Modifying a pre-trained network & 
\cite{Chen2018Detection},
\cite{Gao2018Diagnosis},
\cite{Kassani2019Diabetic},
\cite{Wijesinghe2019Transfer},
\cite{Ahmad2019Deep},
\cite{Islam2019DeepDR},
\cite{Yu2020Performance} \\

Not stated &
\cite{Gulshan2016Development},
\cite{Hazim2018Early},
\cite{Laxmi2020Adaptive} \\

\bottomrule
\end{tabular}%
}
\end{table}

As can be seen from Table~\ref{table-transfer-learning-approaches}, fine-tuning the whole network is the most preferred approach, where using the pre-trained network without fine-tuning is the least preferred one. 
Besides, training the state-of-the-art architecture from scratch is the least preferred approach due to the long training time. 
Additionally, in some of the studies, higher performances were achieved by modifying the state-of-the-art architecture trained by ImageNet.

\section{Diabetic Retinopathy Datasets}
\label{section-datasets}

There are various publicly available datasets, including images for classification or detection of DR. 
Table~\ref{table-dataset-size-link} shows the used datasets among the reviewed \countOfReviewedPapers{} articles.
Properties of datasets, such as their sizes, the number of classes they contain, and their download links, are given. 

\begin{table}[!htb]
\centering
\caption{DR dataset properties and access links.}
\label{table-dataset-size-link}
\resizebox{\textwidth}{!}{%
\begin{tabular}{p{0.3\textwidth}p{0.13\textwidth}p{0.05\textwidth}p{0.95\textwidth}}

\toprule
Name  & Data Size & Class & Link \\
\midrule

DR Detection Kaggle 	& 88702 & 5 & \url{https://www.kaggle.com/c/diabetic-retinopathy-detection/data} \\
Messidor  	& 1200 & 4 & \url{https://www.adcis.net/en/third-party/messidor/} \\
Messidor-2  & 1748 & 5 & \url{https://www.adcis.net/en/third-party/messidor2/} \\
APTOS 2019 	& 3662 & 5 & \url{https://www.kaggle.com/c/aptos2019-blindness-detection} \\
DIARETDB1  	& 89 & 2 & \url{http://www2.it.lut.fi/project/imageret/diaretdb1/} \\
IDRiD  		& 516 & 5 & \url{https://idrid.grand-challenge.org/Data_Download/} \\
eOphtha  	& 463 & 3 & \url{https://www.adcis.net/en/third-party/e-ophtha/} \\
DIABRET  	& 1331 & 5 & \url{https://www.kaggle.com/lrasmy/sample-diab-retn} \\
DR1  		& 1014 & 2 & \url{http://www.record.ic.unicamp.br/site/asdr} \\
University of Auckland Diabetic Retinopathy (UoA-DR)		& 200 & 3 & \url{https://researchspace.auckland.ac.nz/handle/2292/46737} \\
STARE  		& 397 & 14 & \url{https://cecas.clemson.edu/~ahoover/stare/} \\
ODIR 2019 	& 5814 & 8 & \url{https://odir2019.grand-challenge.org/dataset/} \\

\bottomrule

\end{tabular}%
}

\end{table}

\begin{figure}[ht!]
  \includegraphics[scale=0.5]{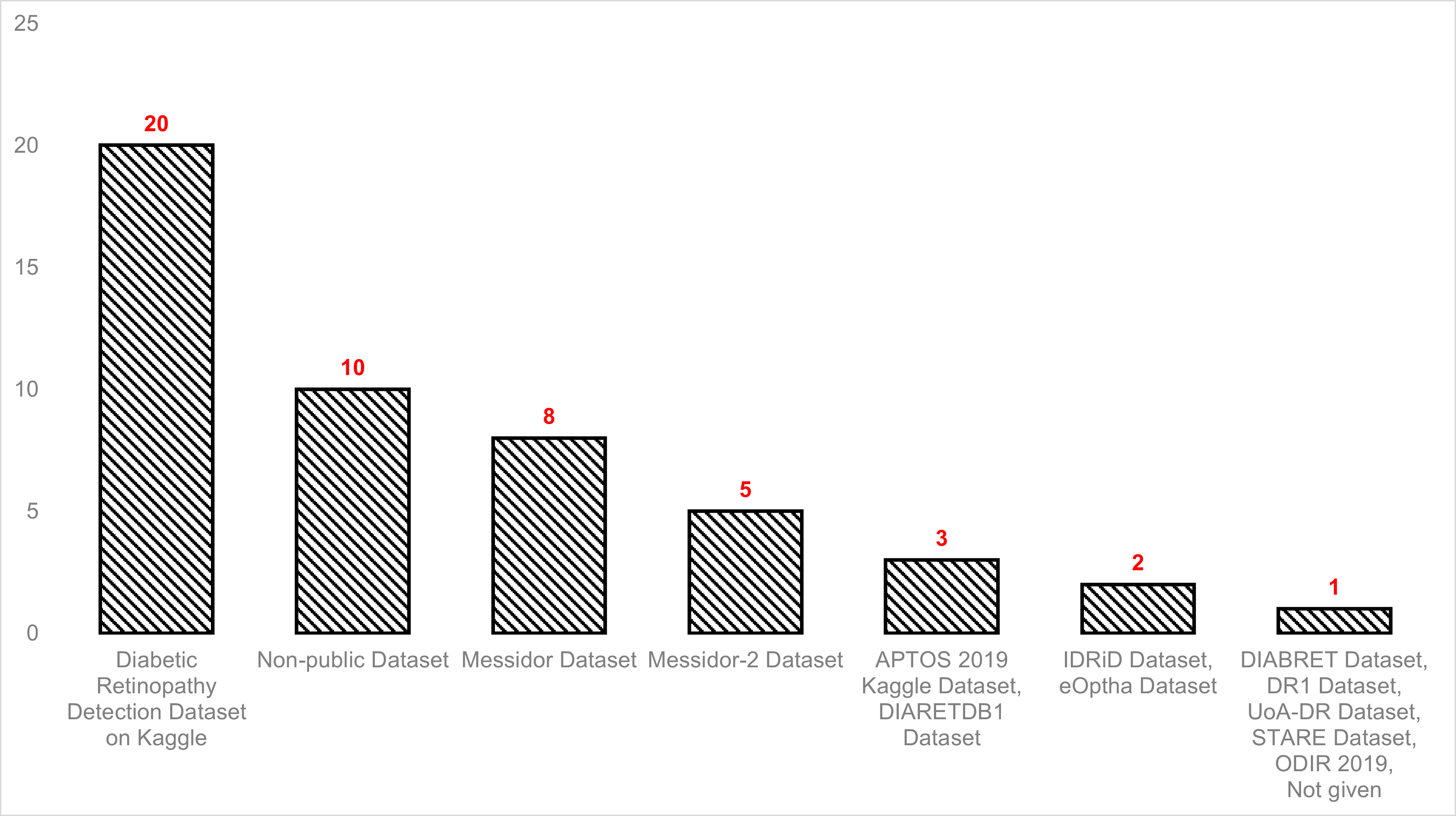}
  \centering
  \caption{The frequencies of the used datasets.}
  \label{figure-dataset-study}
\end{figure}

Table~\ref{table-dataset-study} demonstrates the datasets used in the reviewed articles, and in Figure~\ref{figure-dataset-study} the frequencies of the used datasets are shown. 
As shown in Figure~\ref{figure-dataset-study}, many publicly available DR datasets were utilized for DR classification. 
Among them, the Diabetic Retinopathy Detection Dataset on Kaggle was the most preferred because of its size and accessibility. 
Besides, it is noteworthy that many non-public datasets were also used in most studies \cite{Li2019Automatic,Li2019Diagnostic,Ahmad2019Deep,Yip2019Enhanced,Roshan2020Fine,Takahashi2017Applying, Yu2020Performance,Gao2018Diagnosis,Zhang2019Automated}. 
Deep neural networks work better on large datasets, and the size of the data set is a very important parameter in the network's performance. 
For this reason, most studies have used more than one dataset to improve classification performance \cite{Li2017Convolutional, Zhou2018Deep, Wijesinghe2019Transfer, Ahmad2019Deep, Voets2019Reproduction, Laxmi2020Adaptive, Hacisoftaoglu2020Deep, Gulshan2016Development, Vo2016New, Lam2018Retinal, Lam2018Automated, Mateen2020Exudate, Hathwar2019Automated}. 

\begin{table}[!htb]
\centering
\caption{Data sets used in the reviewed papers.}
\label{table-dataset-study}
\resizebox{\textwidth}{!}{%

\begin{tabular}{p{0.5\textwidth}p{0.07\textwidth}p{0.5\textwidth}}

\toprule
Dataset & Count & Study   \\

\midrule 

Diabetic Retinopathy Detection Dataset on Kaggle &  20 &\cite{Zhou2018Deep},
\cite{Wan2018Deep},
\cite{Wang2018Diabetic},
\cite{Hagos2019Transfer},
\cite{Voets2019Reproduction},
\cite{Mohammadian2017Comparative},
\cite{Lian2018Deep},
\cite{Lam2018Automated},
\cite{Chen2018Detection},
\cite{Lam2018Retinal},
\cite{Vo2016New},
\cite{Wijesinghe2019Transfer},
\cite{Islam2019DeepDR}, 
\cite{Roshan2020Fine},
\cite{Hacisoftaoglu2020Deep},
\cite{Laxmi2020Adaptive},
\cite{Zeng2019Automated},
\cite{Math2019Identification},
\cite{Gulshan2016Development},    
\cite{Hathwar2019Automated} \\
Non-public Dataset  & 10 &
\cite{Li2019Automatic},
\cite{Li2019Diagnostic},
\cite{Ahmad2019Deep},
\cite{Yip2019Enhanced},
\cite{Roshan2020Fine},
\cite{Takahashi2017Applying},
\cite{Yu2020Performance},
\cite{Gao2018Diagnosis},
\cite{Hacisoftaoglu2020Deep},
\cite{Zhang2019Automated} \\
Messidor Dataset  & 8 &
\cite{Zhou2018Deep},
\cite{Li2017Convolutional},
\cite{Hacisoftaoglu2020Deep},
\cite{Vo2016New},
\cite{Hazim2018Early},
\cite{Yu2020Performance},
\cite{Lam2018Automated},
\cite{Qomariah2019Classification}  \\
Messidor-2 Dataset & 5 &
\cite{Voets2019Reproduction},
\cite{Ahmad2019Deep},
\cite{Hacisoftaoglu2020Deep},
\cite{Gulshan2016Development},
\cite{Li2019Automatic} \\
APTOS 2019 Kaggle Dataset & 3 & 
\cite{Bodapati2020Blended},
\cite{Dekhil2019Deep},
\cite{Kassani2019Diabetic} \\
DIARETDB1 Dataset  & 3 &
\cite{Zhou2018Deep},
\cite{Laxmi2020Adaptive},
\cite{Mateen2020Exudate} \\
IDRiD Dataset  & 2 &
\cite{Hacisoftaoglu2020Deep},
\cite{Hathwar2019Automated} \\
eOptha Dataset  & 2 &
\cite{Lam2018Retinal},
\cite{Mateen2020Exudate} \\
DIABRET Dataset & 1 &
\cite{Wijesinghe2019Transfer}  \\
DR1 Dataset  & 1 &
\cite{Li2017Convolutional}  \\
UoA-DR Dataset & 1 &
\cite{Hacisoftaoglu2020Deep}  \\
STARE Dataset  & 1 & 
\cite{Choi2017Multi} \\
ODIR 2019  & 1 &
\cite{Wang2020Multi} \\
Not given  & 1 &
\cite{Xu2018Improved}\\ 
\bottomrule
\end{tabular}%
}
\end{table}                                                                                                                                                                                                                                                                                                                                               

\subsection{Datasets: the number of classes used} 

\begin{figure}[ht!]
  \includegraphics[scale=0.5]{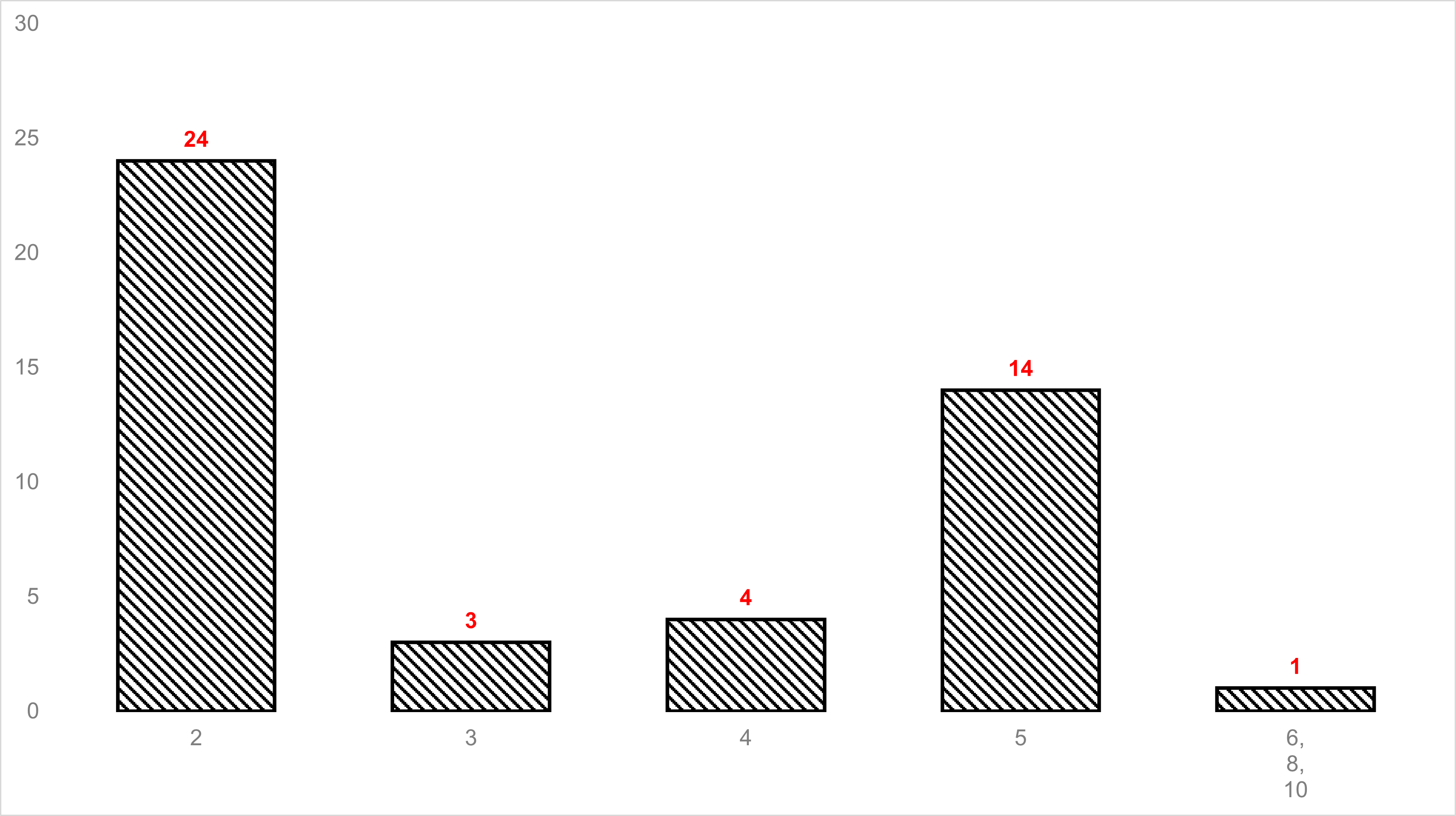}
  \centering
  \caption{The frequencies of number of classes used.}
  \label{figure-study-noofclass}
\end{figure}

Table~\ref{table-study-noofclass} presents the number of classes for DR classification used in the reviewed papers.
Figure~\ref{figure-study-noofclass} shows the frequencies of the used number of classes.
As mentioned before, the number of normal and abnormal images included in each dataset is different. 
In particular, the number of abnormal images graded according to DR severity level is inconsistent. 
The imbalance of the datasets and the quality of the images make it difficult to evaluate for more than two classes. 
For this reason, binary classification was used more than multi-class classification in the many reviewed studies \cite{Gulshan2016Development,Vo2016New,Choi2017Multi,Lam2018Retinal,Mohammadian2017Comparative,Li2017Convolutional,Zhou2018Deep,Hazim2018Early,Lam2018Automated,Hathwar2019Automated,Yip2019Enhanced,Qomariah2019Classification,Hagos2019Transfer,Voets2019Reproduction,Zeng2019Automated,Laxmi2020Adaptive,Mateen2020Exudate,Hacisoftaoglu2020Deep,Roshan2020Fine,Yu2020Performance,Bodapati2020Blended}.

\begin{table}[!htb]
\centering
\caption{Number of classes for DR classification.}
\label{table-study-noofclass}
\resizebox{\textwidth}{!}{%

\begin{tabular}{p{0.1\textwidth}p{0.9\textwidth}}

\toprule
Number of class & Study \\
\midrule

2  &
\cite{Gulshan2016Development},
\cite{Vo2016New},
\cite{Choi2017Multi},
\cite{Lam2018Retinal},
\cite{Mohammadian2017Comparative},
\cite{Li2017Convolutional},
\cite{Zhou2018Deep},
\cite{Hazim2018Early},
\cite{Lam2018Automated},
\cite{Hathwar2019Automated},
\cite{Yip2019Enhanced}, 
\cite{Qomariah2019Classification},
\cite{Hagos2019Transfer},
\cite{Voets2019Reproduction},
\cite{Zeng2019Automated},
\cite{Laxmi2020Adaptive},    
\cite{Mateen2020Exudate},
\cite{Hacisoftaoglu2020Deep},    
\cite{Roshan2020Fine},
\cite{Yu2020Performance},
\cite{Bodapati2020Blended},
\cite{Li2019Automatic},
\cite{Math2019Identification},
\cite{Zhang2019Automated} \\

3  & 
\cite{Lam2018Automated},
\cite{Ahmad2019Deep},
\cite{Choi2017Multi}\\

4  & 
\cite{Takahashi2017Applying},
\cite{Gao2018Diagnosis},
\cite{Lam2018Automated}, 
\cite{Zhang2019Automated} \\

5 & 
\cite{Lam2018Retinal},
\cite{Chen2018Detection},
\cite{Lian2018Deep},
\cite{Wan2018Deep},
\cite{Wang2018Diabetic} ,    
\cite{Xu2018Improved},    
\cite{Kassani2019Diabetic},
\cite{Wijesinghe2019Transfer},    
\cite{Li2019Automatic},    
\cite{Dekhil2019Deep},    
\cite{Islam2019DeepDR},    
\cite{Bodapati2020Blended},    
\cite{Zeng2019Automated}, 
\cite{Vo2016New} \\

6 & 
\cite{Li2019Diagnostic} \\
8 & 
\cite{Wang2020Multi} \\
10 & 
\cite{Choi2017Multi} \\ 
\bottomrule
\end{tabular}%
}
\end{table}

\subsection{Preprocessing and Data Augmentation}
\label{preprocessing-and-dataaugmentation}

Deep learning algorithms are used in order to eliminate the need for handcrafted features and, therefore pre-processing steps. 
However, in some DNN studies, some pre-processing steps for enhancing and improving the image quality are applicable. 
Additionally, several pre-processing techniques for identifying the region of interest (ROI) in images are used commonly.
Applied pre-processing methods in the reviewed studies are presented in Table~\ref{table-preprocessing}. 
Resizing is the most utilized pre-processing step. The basis for this is to make the images appropriate for the input size of CNN. 
Resizing is followed by normalization and cropping, as expected. 
Normalizing is performed to promote the DNN learning process, make the training faster, and avoid overfitting. 
Further, cropping improves success by reducing the contribution of the background to the training process.

\begin{table}[!htb]
\centering
\caption{Applied pre-processsing methods in the review studies.}
\label{table-preprocessing}
\resizebox{\textwidth}{!}{%
\begin{tabular}{p{0.7\textwidth}p{0.3\textwidth}}

\toprule

Pre-Processing Method & Study \\
\midrule

Resize/Reshape/Rescale & 
\cite{Gulshan2016Development},
\cite{Mohammadian2017Comparative},
\cite{Wang2020Multi},
\cite{Li2017Convolutional},
\cite{Chen2018Detection},
\cite{Lian2018Deep},
\cite{Hazim2018Early},
\cite{Wang2018Diabetic},
\cite{Hathwar2019Automated},
\cite{Kassani2019Diabetic},
\cite{Wijesinghe2019Transfer},
\cite{Li2019Diagnostic},
\cite{Qomariah2019Classification},
\cite{Hagos2019Transfer},
\cite{Voets2019Reproduction},
\cite{Zeng2019Automated},
\cite{Bodapati2020Blended},
\cite{Laxmi2020Adaptive},
\cite{Roshan2020Fine},
\cite{Hacisoftaoglu2020Deep},
\cite{Islam2019DeepDR},
\cite{Li2019Automatic},
\cite{Choi2017Multi},
\cite{Ahmad2019Deep},
\cite{Choi2017Multi},
\cite{Mateen2020Exudate},
\cite{Vo2016New},
\cite{Zhou2018Deep},
\cite{Math2019Identification}
\\

Normalization & 
\cite{Gulshan2016Development},
\cite{Lam2018Retinal},
\cite{Zhou2018Deep},
\cite{Lian2018Deep},
\cite{Gao2018Diagnosis},
\cite{Lam2018Automated},
\cite{Wan2018Deep},
\cite{Kassani2019Diabetic},
\cite{Islam2019DeepDR},
\cite{Zeng2019Automated},
\cite{Yu2020Performance},
\cite{Laxmi2020Adaptive}\\

Crop & 
\cite{Vo2016New},
\cite{Lam2018Retinal},
\cite{Wang2020Multi},
\cite{Li2017Convolutional},
\cite{Chen2018Detection} 
\cite{Lam2018Automated},
\cite{Dekhil2019Deep},
\cite{Qomariah2019Classification},
\cite{Hagos2019Transfer},
\cite{Yu2020Performance},
\cite{Hacisoftaoglu2020Deep},
\cite{Zhou2018Deep},
\cite{Wan2018Deep},
\cite{Math2019Identification}\\

Remove an average image from each of the images, local average magnitude subtraction, Subtract each pixel value of images by the weighted means  of its surrounding pixel values, and add it by 50 \%  grayscale* & 
\cite{Li2017Convolutional},
\cite{Hathwar2019Automated},
\cite{Hagos2019Transfer},
\cite{Zeng2019Automated} *, 
\cite{Roshan2020Fine} * \\

Contrast adjustment/enhancement/improvement & 

\cite{Lam2018Automated},
\cite{Li2019Automatic},
\cite{Islam2019DeepDR},
\cite{Laxmi2020Adaptive},
\cite{Yu2020Performance},
\cite{Zhou2018Deep},
\cite{Zhang2019Automated} \\

Not given & 
\cite{Xu2018Improved},
\cite{Yip2019Enhanced},
\cite{Takahashi2017Applying}\\

Scaling & 
\cite{Chen2018Detection},
\cite{Math2019Identification}\\

Clipped & 
\cite{Vo2016New},
\cite{Mohammadian2017Comparative},
\cite{Roshan2020Fine} \\

Removal of unwanted background, cropped out all backgrounds & 
\cite{Wijesinghe2019Transfer},
\cite{Li2019Diagnostic},
\cite{Lian2018Deep},
\cite{Zeng2019Automated},
\cite{Zhang2019Automated}\\

Downsampling & 
\cite{Dekhil2019Deep} ,  
\cite{Hacisoftaoglu2020Deep} \\

Size standardization & 
\cite{Mateen2020Exudate} \\

Mask & 
\cite{Hacisoftaoglu2020Deep},
\cite{Zhou2018Deep}\\

Color enhancement & 
\cite{Lian2018Deep} \\

Upsampling & 
\cite{Lam2018Retinal} \\

Pixel Mapping & 
\cite{Mohammadian2017Comparative} \\

Sharpness unification & 
\cite{Mohammadian2017Comparative} \\

Histogram Equilization & 
\cite{Wang2020Multi},
\cite{Zhang2019Automated} \\

Min-pooling & 
\cite{Kassani2019Diabetic} \\

Downsize & 
\cite{Li2019Automatic} \\

Nonlocal means denoising & 
\cite{Li2019Automatic},
\cite{Wan2018Deep}\\

Resampling & 
\cite{Li2019Diagnostic} \\

Unsharp masking & 
\cite{Islam2019DeepDR} \\

Brightness adjustment & 
\cite{Yu2020Performance} \\

Noise reduction & 
\cite{Yu2020Performance} \\ 

Oversampling & 
\cite{Choi2017Multi} \\

Illumination &
\cite{Laxmi2020Adaptive} \\

Illumination equalization &
\cite{Zhou2018Deep} \\

Gaussian smoothing &
\cite{Math2019Identification} \\

Gamma correction &
\cite{Ahmad2019Deep} \\

\bottomrule
\end{tabular}%
}
\end{table}

Data augmentation is a commonly performed technique in deep learning applications to reduce overfitting due to the lack of training data set and increase the algorithm's performance by increasing the amount of data. 
Traditional data augmentation techniques are translation, stretching, flipping, zooming, contrast adjustment, and rotation.
As shown in Table~\ref{table-augmentation},  in most of the studies,  traditional data augmentation techniques  are used, however, in a minority of the studies, different approaches are used. 
At the same time, as can be seen from Table~\ref{table-augmentation} in almost half of the studies data augmentation process is not mentioned nor the used technique is explained. 
Nevertheless, the effect of data augmentation in DR classification is still a question mark and needs further investigation.

\begin{table}[!htb]
\centering
\caption{Applied data augmentation methods in the reviewed studies.}
\label{table-augmentation}
\resizebox{\textwidth}{!}{%
\begin{tabular}{p{0.4\textwidth}p{0.6\textwidth}}

\toprule

Data Augmentation Method & Study \\
\midrule

Not given & 
\cite{Gulshan2016Development},
\cite{Vo2016New},
\cite{Takahashi2017Applying},
\cite{Lian2018Deep},
\cite{Hazim2018Early},
\cite{Wang2018Diabetic},
\cite{Wijesinghe2019Transfer},
\cite{Li2019Automatic},
\cite{Math2019Identification},
\cite{Yip2019Enhanced},
\cite{Dekhil2019Deep},
\cite{Qomariah2019Classification},
\cite{Hagos2019Transfer},
\cite{Voets2019Reproduction},
\cite{Bodapati2020Blended},
\cite{Laxmi2020Adaptive},
\cite{Mateen2020Exudate}\\

Rotation & 
\cite{Choi2017Multi},
\cite{Mohammadian2017Comparative},
\cite{Wang2020Multi},
\cite{Li2017Convolutional},
\cite{Zhou2018Deep},
\cite{Chen2018Detection},
\cite{Gao2018Diagnosis},
\cite{Lam2018Automated},
\cite{Wan2018Deep},
\cite{Xu2018Improved},
\cite{Hathwar2019Automated},
\cite{Li2019Diagnostic},
\cite{Ahmad2019Deep},
\cite{Islam2019DeepDR},
\cite{Yu2020Performance},
\cite{Roshan2020Fine},
\cite{Zeng2019Automated},
\cite{Zhang2019Automated} \\

Flipping & 
\cite{Mohammadian2017Comparative},
\cite{Li2017Convolutional},
\cite{Zhou2018Deep},
\cite{Chen2018Detection},
\cite{Gao2018Diagnosis},
\cite{Wan2018Deep},
\cite{Hathwar2019Automated},
\cite{Li2019Diagnostic},
\cite{Ahmad2019Deep},
\cite{Islam2019DeepDR},
\cite{Hacisoftaoglu2020Deep},
\cite{Roshan2020Fine},
\cite{Zeng2019Automated},
\cite{Zhang2019Automated} \\

Translation & 
\cite{Choi2017Multi},
\cite{Lam2018Retinal},
\cite{Wang2020Multi},
\cite{Zhou2018Deep},
\cite{Li2019Diagnostic},
\cite{Zeng2019Automated} \\

Zoom-in and zoom-out & 
\cite{Lam2018Retinal},
\cite{Gao2018Diagnosis},
\cite{Lam2018Automated},
\cite{Hathwar2019Automated},
\cite{Ahmad2019Deep} \\

Brightness & 
\cite{Choi2017Multi}, 
\cite{Lam2018Retinal},
\cite{Islam2019DeepDR} \\

Contrast & 
\cite{Lam2018Retinal},
\cite{Chen2018Detection}
\cite{Islam2019DeepDR} \\

Stretching & 
\cite{Wan2018Deep},
\cite{Xu2018Improved},
\cite{Li2019Diagnostic} \\

Rolling of the image & 
\cite{Lam2018Retinal},
\cite{Lam2018Automated} \\

Shifting & 
\cite{Mohammadian2017Comparative},    
\cite{Roshan2020Fine} \\

Scaling & 
\cite{Zhou2018Deep},
\cite{Zeng2019Automated},
\cite{Chen2018Detection}\\

Cropped & 
\cite{Chen2018Detection},
\cite{Zeng2019Automated} \\

Shearing & 
\cite{Hathwar2019Automated},    
\cite{Zeng2019Automated}  \\

Additive Gaussian noise & 
\cite{Choi2017Multi} \\

Distort & 
\cite{Gao2018Diagnosis} \\

Padding & 
\cite{Lam2018Automated} \\

Color augmentation & 
\cite{Li2019Diagnostic} \\

Saturation & 
\cite{Islam2019DeepDR} \\

Shuffle & 
\cite{Islam2019DeepDR} \\

Mirror & 
\cite{Yu2020Performance},
\cite{Hacisoftaoglu2020Deep} \\

Inverting & 
\cite{Zeng2019Automated} \\ 

Not applied &
\cite{Kassani2019Diabetic} \\

\bottomrule
\end{tabular}%
}
\end{table}

\section{Pre-trained CNN Architectures}
\label{architectures}

A pre-trained CNN model is a model which uses parameters of a model that is trained on another problem to solve a similar task rather than training a new model from scratch. 
Many of the popular pre-trained CNN models are trained using ImageNet \cite{Kieu2020Suvey}. 
Generally, though some topological differences are presented in modern architectures, the basic components (e.g., convolutional and pooling layers) of the models are almost identical \cite{Alom2019State}. 
Accordingly, since the introduction of AlexNet many highly thriving architectures, such as VGG, ResNet, DenseNet, GoogleNet, were presented for image classification. \cite{Kandel2020Transfer}. 
Therefore, many of these well-known pre-trained CNN architectures have also been used for DR detection.


In most of the studies reviewed, different architectures were compared to achieve the best performance \cite{Wan2018Deep, Wang2018Diabetic, Li2019Diagnostic, Bodapati2020Blended, Mohammadian2017Comparative, Li2017Convolutional, Lian2018Deep, Wijesinghe2019Transfer, Ahmad2019Deep, Yip2019Enhanced, Islam2019DeepDR, Hacisoftaoglu2020Deep, Vo2016New, Choi2017Multi, Lam2018Retinal, Wang2020Multi, Gao2018Diagnosis, Lam2018Automated, Yu2020Performance, Mateen2020Exudate, Qomariah2019Classification, Kassani2019Diabetic, Hathwar2019Automated, Zhang2019Automated}. 
Whereas, only one architecture is analyzed in some of the studies \cite{Zhou2018Deep, Li2019Automatic, Voets2019Reproduction,  Laxmi2020Adaptive, Roshan2020Fine, Gulshan2016Development, Takahashi2017Applying, Chen2018Detection, Hazim2018Early, Zeng2019Automated, Hagos2019Transfer, Dekhil2019Deep, Math2019Identification, Xu2018Improved}.

\begin{figure}[ht!]
  \includegraphics[scale=0.35]{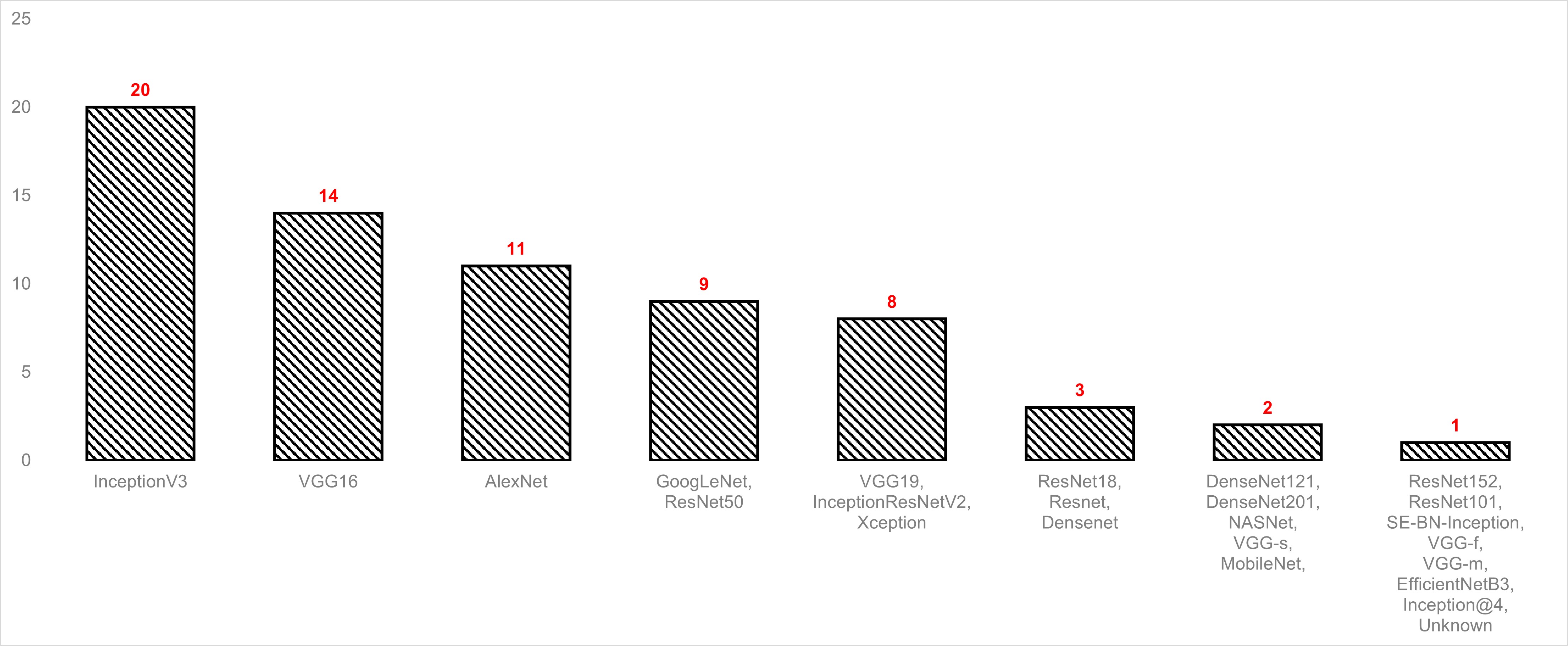}
  \centering
  \caption{The frequencies of the used architectures.}
  \label{figure-architecture-study}
\end{figure}

Table~\ref{table-architectures} illustrates the architectures used in the reviewed studies, and Figure~\ref{figure-architecture-study} shows the frequencies of the used architectures.
As can be seen from Table~\ref{table-architectures}, Inception-v3 is the most commonly used architecture (14 studies) among the many state-of-the-art architectures in the literature. 
The selection of Inception-V3 architecture may be due to the obtained optimized results using this architecture \cite{Islam2019DeepDR}.
Subsequent to Inception-v3, VGG-16 and AlexNet are the architectures that are mostly preferred in the reviewed articles.

In the studies that investigated the performances of different architectures, AlexNet produced the lowest performances most of the time \cite{Wan2018Deep, Wang2018Diabetic, Li2017Convolutional, Lian2018Deep, Hacisoftaoglu2020Deep, Choi2017Multi, Lam2018Retinal, Lam2018Automated, Qomariah2019Classification}. 
According to \cite{Kandel2020Transfer}, this situation occurs since AlexNet is a shallow model, and the sufficient number of convolutional layers makes the classification a challenge. 
The architecture that yields the best performance among the investigated studies is Inception-V3 \cite{Wang2018Diabetic, Mohammadian2017Comparative, Islam2019DeepDR, Lam2018Retinal, Qomariah2019Classification}.  
The reason of this performance may lie behind the inception modules that are used by Inception-V3. 
Inception modules are consist of different sized filters on the same convolution layer; hence they extract features in different aspect ratios. 
Since features of DR differ in size, the inception module can be beneficial for DR classification \cite{Hagos2019Transfer}.

\begin{table}[!htb]
\centering
\caption{CNN architectures used in the review studies.}
\label{table-architectures}
\resizebox{\textwidth}{!}{%

\begin{tabular}{p{0.3\textwidth}p{0.7\textwidth}}
\toprule
CNN architecture  & Study \\
\midrule
InceptionV3     &

\cite{Wang2018Diabetic},
\cite{Li2019Automatic}, 
\cite{Mohammadian2017Comparative},
\cite{Voets2019Reproduction}, 
\cite{Islam2019DeepDR},
\cite{Ahmad2019Deep},
\cite{Roshan2020Fine}, 
\cite{Gulshan2016Development}, 
\cite{Lam2018Retinal}, 
\cite{Wang2020Multi}, 
\cite{Chen2018Detection}, 
\cite{Gao2018Diagnosis}, 
\cite{Yu2020Performance}, 
\cite{Mateen2020Exudate},  
\cite{Zeng2019Automated}, 
\cite{Hagos2019Transfer}, 
\cite{Qomariah2019Classification}, 
\cite{Kassani2019Diabetic},
\cite{Wijesinghe2019Transfer},
\cite{Zhang2019Automated} \\

VGG16 & 
\cite{Wan2018Deep}, 
\cite{Wang2018Diabetic}, 
\cite{Bodapati2020Blended}, 
\cite{Li2017Convolutional}, 
\cite{Lian2018Deep} , 
\cite{Wijesinghe2019Transfer}, 
\cite{Ahmad2019Deep}, 
\cite{Yip2019Enhanced}, 
\cite{Vo2016New}, 
\cite{Lam2018Retinal}, 
\cite{Wang2020Multi}, 
\cite{Qomariah2019Classification},
\cite{Li2019Diagnostic},
\cite{Dekhil2019Deep}\\

AlexNet & 
\cite{Zhou2018Deep}, 
\cite{Wan2018Deep}, 
\cite{Li2017Convolutional}, 
\cite{Wang2018Diabetic}, 
\cite{Lian2018Deep}, 
\cite{Hacisoftaoglu2020Deep}, 
\cite{Choi2017Multi}, 
\cite{Lam2018Retinal}, 
\cite{Hazim2018Early}, 
\cite{Lam2018Automated},
\cite{Qomariah2019Classification},    
\\

GoogLeNet&  
\cite{Wan2018Deep},
\cite{Li2017Convolutional}, 
\cite{Hacisoftaoglu2020Deep}, 
\cite{Vo2016New} ,
\cite{Lam2018Retinal}, 
\cite{Lam2018Automated}, 
\cite{Qomariah2019Classification},
\cite{Li2019Diagnostic},
\cite{Takahashi2017Applying}\\

VGG19 & 
\cite{Wan2018Deep}, 
\cite{Li2017Convolutional}, 
\cite{Ahmad2019Deep}, 
\cite{Choi2017Multi}, 
\cite{Wang2020Multi}, 
\cite{Gao2018Diagnosis}, 
\cite{Mateen2020Exudate}, 
\cite{Qomariah2019Classification} \\

ResNet50 & 
\cite{Lian2018Deep}, 
\cite{Ahmad2019Deep}, 
\cite{Yip2019Enhanced} 
\cite{Hacisoftaoglu2020Deep}, 
\cite{Wang2020Multi}, 
\cite{Mateen2020Exudate}, 
\cite{Qomariah2019Classification},  
\cite{Kassani2019Diabetic},
\cite{Zhang2019Automated} \\

InceptionResNetV2 & 
\cite{Ahmad2019Deep}, 
\cite{Bodapati2020Blended}, 
\cite{Wang2020Multi}, 
\cite{Yu2020Performance}, 
\cite{Qomariah2019Classification}, 
\cite{Hathwar2019Automated},
\cite{Wijesinghe2019Transfer},
\cite{Zhang2019Automated} \\

Xception & 
\cite{Bodapati2020Blended}, 
\cite{Mohammadian2017Comparative}, 
\cite{Ahmad2019Deep}, 
\cite{Wang2020Multi}, 
\cite{Kassani2019Diabetic}, 
\cite{Hathwar2019Automated},
\cite{Wijesinghe2019Transfer},
\cite{Zhang2019Automated}\\

DenseNet121 & 
\cite{Li2019Diagnostic}, 
\cite{Ahmad2019Deep} \\

DenseNet201  & 
\cite{Wijesinghe2019Transfer}, 
\cite{Qomariah2019Classification} \\

ResNet18 & 
\cite{Li2019Diagnostic}, 
\cite{Wijesinghe2019Transfer}, 
\cite{Gao2018Diagnosis} \\

NASNet  & 
\cite{Ahmad2019Deep}, 
\cite{Bodapati2020Blended} \\

VGG-s & 
\cite{Wan2018Deep}, 
\cite{Li2017Convolutional} \\

MobileNet & 
\cite{Wang2020Multi}, 
\cite{Kassani2019Diabetic}   \\

Resnet & 
\cite{Lam2018Retinal},
\cite{Takahashi2017Applying},
\cite{Wan2018Deep}\\

Densenet & 
\cite{Wang2020Multi},   
\cite{Xu2018Improved},
\cite{Zhang2019Automated}\\

ResNet101 &
 \cite{Gao2018Diagnosis} \\
 
SE-BN-Inception & 
\cite{Li2019Automatic} \\

VGG-f & 
\cite{Li2017Convolutional} \\

VGG-m & 
\cite{Li2017Convolutional}  \\

EfficientNetB3 & 
\cite{Wang2020Multi} \\

Inception@4 & 
\cite{Gao2018Diagnosis} \\

Unknown  & 
\cite{Math2019Identification} \\  

\bottomrule
\end{tabular}%
}
\end{table}     

\section{Performance Metrics}
\label{section-performance-metrics}

\begin{figure}[ht!]
  \includegraphics[scale=0.45]{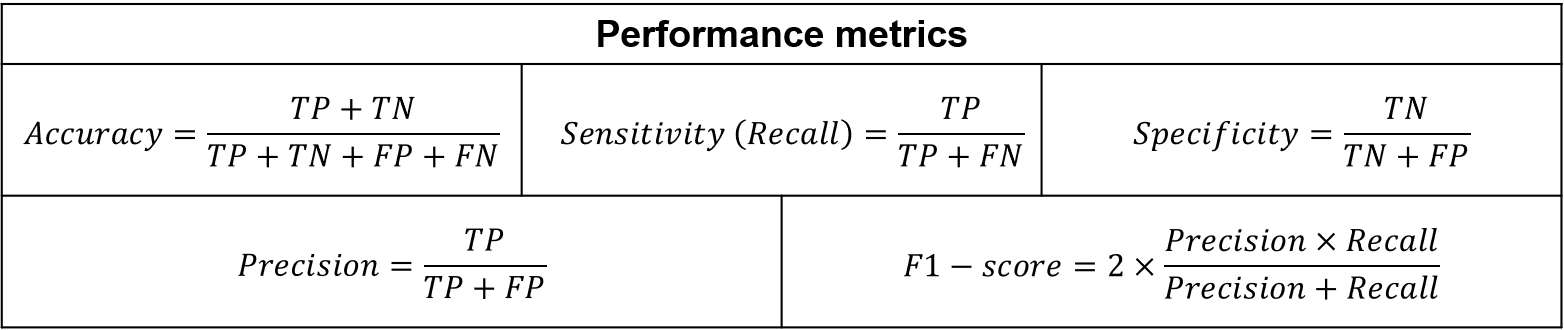}
  \centering
  \caption{Common performance metrics.}
  \label{figure-performance-metrics}
\end{figure}

Common metrics for measuring the performance of classification and detection include accuracy, sensitivity (recall), specificity, precision, F-score, ROC curve.
Similar to other classification problems, common performance metrics used for DR detection. 
The formulation of these metrics has been given in Figure~\ref{figure-performance-metrics}.

Table~\ref{table-performancemetrics-study} provides the performance metrics used in the reviewed studies.
As stated above, accuracy, precision, recall, AUC metrics are typically used to measure the performance of medical classification problems.   
Accordingly, in the reviewed studies, accuracy is the most commonly used metric, followed by sensitivity/recall, AUC, and specificity. 
Moreover, in some of the mentioned studies, \cite{Bodapati2020Blended, Hagos2019Transfer, Islam2019DeepDR, Chen2018Detection, Hacisoftaoglu2020Deep, Li2019Diagnostic, Xu2018Improved, Choi2017Multi, Lam2018Retinal, Zhang2019Automated} metrics that are not so common are used, which inhibits the performance comparison between similar studies. Additionally, time is a significant metric for deep learning structures; however, only one study \cite{Islam2019DeepDR} presents this metric.

\begin{table}[!htb]
\centering
\caption{Performance metrics used in the review studies.}
\label{table-performancemetrics-study}
\resizebox{\textwidth}{!}{%

\begin{tabular}{p{0.5\textwidth}p{0.5\textwidth}}

\toprule

Performance metric & Study \\
\midrule

Accuracy &  

\cite{Wan2018Deep},
\cite{Wang2018Diabetic},
\cite{Lam2018Automated},
\cite{Gao2018Diagnosis},
\cite{Hazim2018Early},
\cite{Takahashi2017Applying},
\cite{Choi2017Multi},
\cite{Vo2016New},
\cite{Kassani2019Diabetic},
\cite{Math2019Identification},
\cite{Dekhil2019Deep},
\cite{Qomariah2019Classification}, 
\cite{Hagos2019Transfer},
\cite{Mateen2020Exudate},
\cite{Yu2020Performance},
\cite{Li2019Automatic},
\cite{Li2019Diagnostic},
\cite{Bodapati2020Blended},
\cite{Mohammadian2017Comparative},
\cite{Roshan2020Fine},
\cite{Hacisoftaoglu2020Deep},
\cite{Ahmad2019Deep},
\cite{Wijesinghe2019Transfer},
\cite{Lian2018Deep},
\cite{Li2017Convolutional},
\cite{Yu2020Performance},
\cite{Wang2020Multi},
\cite{Islam2019DeepDR},
\cite{Zhang2019Automated}
\\

Sensivity/Recall & 

\cite{Zhou2018Deep} 
\cite{Wan2018Deep},
\cite{Lam2018Automated},
\cite{Vo2016New},
\cite{Gulshan2016Development},
\cite{Hathwar2019Automated},
\cite{Kassani2019Diabetic},
\cite{Yu2020Performance},
\cite{Li2019Automatic},
\cite{Voets2019Reproduction},
\cite{Li2017Convolutional},
\cite{Zeng2019Automated},
\cite{Yip2019Enhanced},
\cite{Laxmi2020Adaptive},
\cite{Hacisoftaoglu2020Deep},
\cite{Bodapati2020Blended},
\cite{Gao2018Diagnosis},
\cite{Chen2018Detection},
\cite{Mateen2020Exudate},
\cite{Ahmad2019Deep},
\cite{Roshan2020Fine},
\cite{Lam2018Retinal},
\cite{Islam2019DeepDR},
\cite{Wang2020Multi},
\cite{Zhang2019Automated}
\\

AUC & 
\cite{Zhou2018Deep} 
\cite{Wan2018Deep},    
\cite{Wang2020Multi},    
\cite{Lam2018Retinal},    
\cite{Vo2016New},    
\cite{Gulshan2016Development},    
\cite{Kassani2019Diabetic},    
\cite{Math2019Identification},    
\cite{Zeng2019Automated}  
\cite{Yu2020Performance}, 
\cite{Li2019Automatic}, 
\cite{Voets2019Reproduction},
\cite{Li2017Convolutional}, 
\cite{Laxmi2020Adaptive}, 
\cite{Hacisoftaoglu2020Deep}, 
\cite{Yip2019Enhanced},
\cite{Lian2018Deep},
\cite{Choi2017Multi},
\cite{Zhang2019Automated}
\\

Specificity & 
\cite{Wan2018Deep} 
\cite{Li2019Automatic},    
\cite{Lam2018Automated},    
\cite{Vo2016New},    
\cite{Gulshan2016Development},    
\cite{Hathwar2019Automated},    
\cite{Kassani2019Diabetic},    
\cite{Zeng2019Automated},    
\cite{Yu2020Performance}, 
\cite{Voets2019Reproduction},
\cite{Li2017Convolutional}, 
\cite{Yip2019Enhanced},
\cite{Hacisoftaoglu2020Deep}, 
\cite{Laxmi2020Adaptive},
\cite{Zhang2019Automated}
\\

F1-score & 
\cite{Zhou2018Deep},
\cite{Chen2018Detection},
\cite{Wang2020Multi},
\cite{Mateen2020Exudate},
\cite{Bodapati2020Blended},
\cite{Lian2018Deep},
\cite{Ahmad2019Deep},
\cite{Roshan2020Fine},
\cite{Wijesinghe2019Transfer},
\cite{Islam2019DeepDR},
\cite{Zhang2019Automated} \\

Kappa Statistic (Kappa score/Weighted kappa score/ Quadratic weighted kappa/The prevalence- and bias-adjusted kappa) & 
\cite{Bodapati2020Blended},    
\cite{Wang2020Multi},    
\cite{Choi2017Multi},    
\cite{Hathwar2019Automated},    
\cite{Dekhil2019Deep},    
\cite{Zeng2019Automated},
\cite{Takahashi2017Applying},
\cite{Li2019Diagnostic},
\cite{Chen2018Detection},
\cite{Zhang2019Automated}\\

Precision & 
\cite{Zhou2018Deep}, 
\cite{Li2019Diagnostic},
\cite{Gao2018Diagnosis},
\cite{Chen2018Detection},    
\cite{Mateen2020Exudate}, 
\cite{Ahmad2019Deep}, 
\cite{Roshan2020Fine},
\cite{Lam2018Retinal},
\cite{Islam2019DeepDR},
\cite{Bodapati2020Blended},
\cite{Wang2020Multi},
\cite{Zhang2019Automated}\\

Loss  & 
\cite{Bodapati2020Blended},    
\cite{Hagos2019Transfer},
\cite{Islam2019DeepDR},
\cite{Zhang2019Automated}\\

Support & 
\cite{Islam2019DeepDR},   
\cite{Chen2018Detection},
\cite{Zhang2019Automated}\\

Equal Error Rate & 
\cite{Hacisoftaoglu2020Deep}  \\

Intersection Over Union (IoU) & 
\cite{Li2019Diagnostic} \\

Time & 
\cite{Islam2019DeepDR} \\

Error rate  & 
\cite{Xu2018Improved}  \\

Relative classifier information (RCI) & 
\cite{Choi2017Multi}  \\

Area under the precision recall curve & 
\cite{Lam2018Retinal} \\ 

Final score & 
\cite{Wang2020Multi} \\ 

Youden's index &
\cite{Zhang2019Automated} \\

\bottomrule

\end{tabular}
}
\end{table}
\begin{figure}[ht!]
  \includegraphics[scale=0.35]{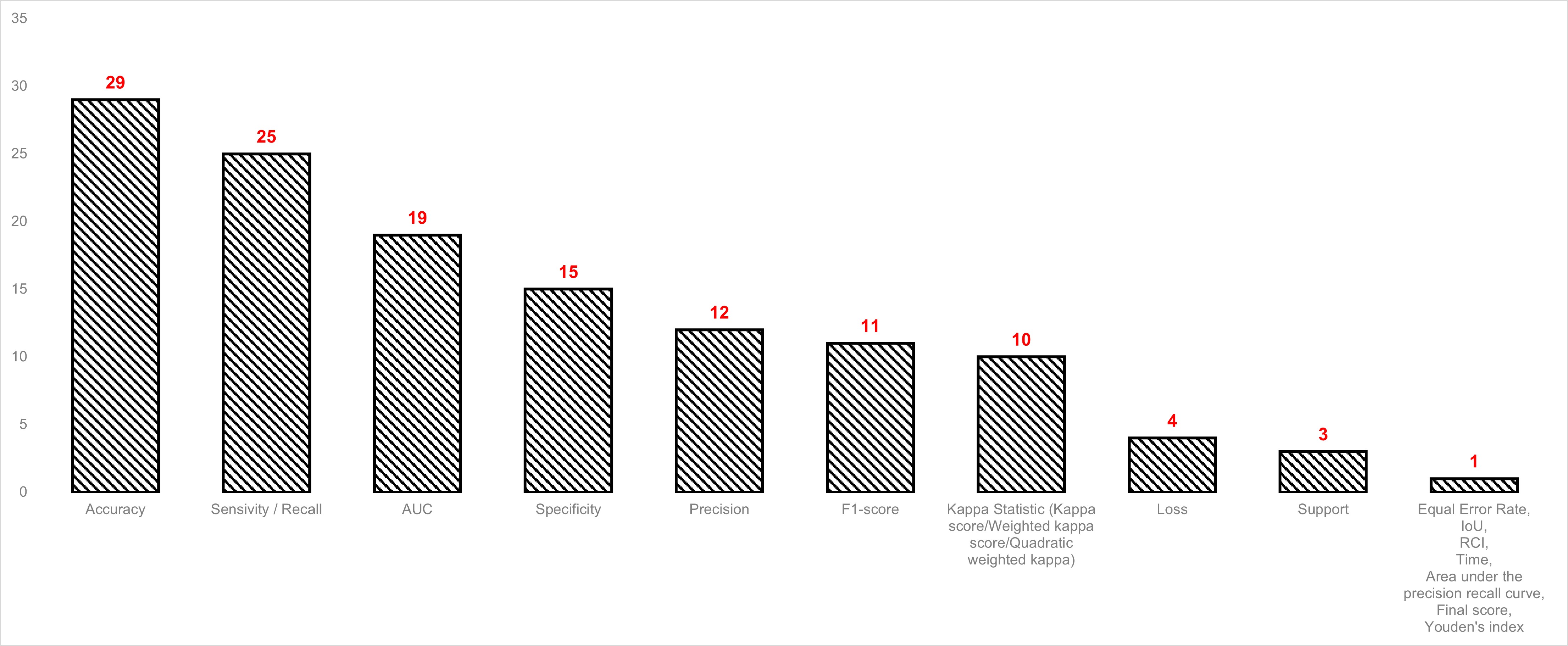}
  \centering
  \caption{The frequencies of the used performance metrics.}
  \label{figure-performancemetrics-study}
\end{figure}

As stated above, accuracy and AUC are the most preferred performance metrics in the reviewed articles. 
Therefore, these two metrics are selected in order to compare the performances of the studies according to data sets. 
Table~\ref{table-dataset-acc-auc} provides the best performance results achieved in the studies according to used datasets.
As seen in the table, Using Diabetic Retinopathy Detection Dataset on Kaggle, the most preferred data set, \cite{Wijesinghe2019Transfer} achieved the highest accuracy whereas the highest AUC is achieved by \cite{Lam2018Retinal}. 
In the Messidor dataset, which is the second most used set, \cite{Hacisoftaoglu2020Deep} obtained the highest accuracy of 99.1\% yet, the highest AUC of 0,98 is obtained by \cite{Li2017Convolutional}. Besides, \cite{Bodapati2020Blended} achieved the highest accuracy using the APTOS data set.


\begin{table}[!htb]
\centering
\caption{Accuracy and AUC}
\label{table-dataset-acc-auc}
\resizebox{\linewidth}{!}{%

\begin{tabular}{p{0.05\textwidth}p{0.1\textwidth}p{0.15\textwidth}p{0.15\textwidth}p{0.05\textwidth}p{0.1\textwidth}p{0.15\textwidth}p{0.15\textwidth}p{0.05\textwidth}p{0.1\textwidth}p{0.15\textwidth}p{0.15\textwidth}}
\toprule
 & Study & Accuracy (\%) & AUC &  & Study & Accuracy (\%) & AUC &  & Study & Accuracy (\%) & AUC \\
\midrule

1&\cite{Dekhil2019Deep}   & 77 & -  & 17& \cite{Math2019Identification}     & 98.46        & 0.985 & 33& \cite{Voets2019Reproduction}      & -            & 0.853  \\
2&\cite{Kassani2019Diabetic}  & 83.09  & -  & 18& \cite{Hacisoftaoglu2020Deep}      & 92.1         & - & 34& \cite{Gulshan2016Development}     & -            & 0.99   \\
3&\cite{Bodapati2020Blended}        & 97.41        & -  &19& \cite{Laxmi2020Adaptive}          & -            & 0.963  & 35& \cite{Ahmad2019Deep}              & 95           & -      \\

4& \cite{Zhou2018Deep}               & -            & 0.925  & 20& \cite{Gulshan2016Development}     & -            &0.991 & 36 & \cite{Li2019Automatic}            & 93.49        & 0.9905 \\
5& \cite{Wan2018Deep}                & 95.68        & 0.9786 &21& \cite{Wijesinghe2019Transfer}     & 98.69        & - & 37 & \cite{Li2019Diagnostic}           & 82.84        & -      \\
6& \cite{Hagos2019Transfer}          & 90.9         & -      &22&\cite{Mateen2020Exudate}          & 98.91        & -  & 38 & \cite{Yip2019Enhanced}            & -            & 0.987  \\
7& \cite{Chen2018Detection}          & 80           & -      &23&\cite{Zhou2018Deep}               & -            & -  & 39 & \cite{Takahashi2017Applying}      & 96           & -      \\
8& \cite{Lam2018Retinal}             & 98           & 0.99   &24&\cite{Li2017Convolutional}        & 94.12        & 0.9786 & 40 & \cite{Yu2020Performance}          & 97.07        & 0.997  \\
9& \cite{Vo2016New}                  & 83.63        & -      &25& \cite{Hathwar2019Automated}       & -            & -  & 41 & \cite{Gao2018Diagnosis}           & 88.72        & -      \\
10& \cite{Wang2018Diabetic}           & 63.23        & -      &26& \cite{Hazim2018Early}             & 88.3         & - & 42 & \cite{Roshan2020Fine}             & 70.30        & -      \\
11& \cite{Voets2019Reproduction}      & -            & 0.951  &27& \cite{Vo2016New}                  & 89.7         & 0.891  &43 & \cite{Hacisoftaoglu2020Deep}      & 98.6         & 0.998  \\
12& \cite{Mohammadian2017Comparative}    & 87.12        & -      &28& \cite{Yu2020Performance}          & 87.18        & 0.926  &44 & \cite{Zhang2019Automated} & 97.67 & 0.986   \\

13& \cite{Lian2018Deep}               & 79.04        & 0.82   &29& \cite{Qomariah2019Classification} & 95.83        & -    &45 & \cite{Choi2017Multi}              & 80.8         & 0.903      \\
14& \cite{Wijesinghe2019Transfer}     & 98.63        & -      &30& \cite{Zhou2018Deep}               & -            & 0.960  &46 & \cite{Mateen2020Exudate}          & 98.43        & -     \\
15& \cite{Zeng2019Automated}          & -            & 0.951  &31& \cite{Li2017Convolutional}        & 92.01        & 0.9834 &47 & \cite{Lam2018Retinal}             & -            & 0.95  \\
16& \cite{Islam2019DeepDR}            & 94.3         & -      &32& \cite{Hacisoftaoglu2020Deep}      & 99.1         & -  & 48 & \cite{Wang2020Multi}              & 89           & 0.73       \\     
                         
 & & & & & & & &49  &  \cite{Lam2018Automated}           & 74.5         & - \\

\end{tabular}%

}

\end{table}
%

\section{Parameter Optimization for Pre-trained CNN}
\label{optimization}

The learning process in multilayer artificial neural networks is an optimization problem. 
During the learning process, many classical and intelligent approaches can be applied. 
Additionally,  some hyper parameter tuning can be included in the learning error minimization process.
For example, using adaptive learning rate or including moment parameter in weight update formulation helps convergence and reduces the training time. 
All of these approaches can be evaluated as hyper parameter tuning.  
According to this subject, some of the evaluated papers, during the learning process, used adaptive learning rate and momentum parameter.

\begin{figure}[ht!]
  \includegraphics[scale=0.5]{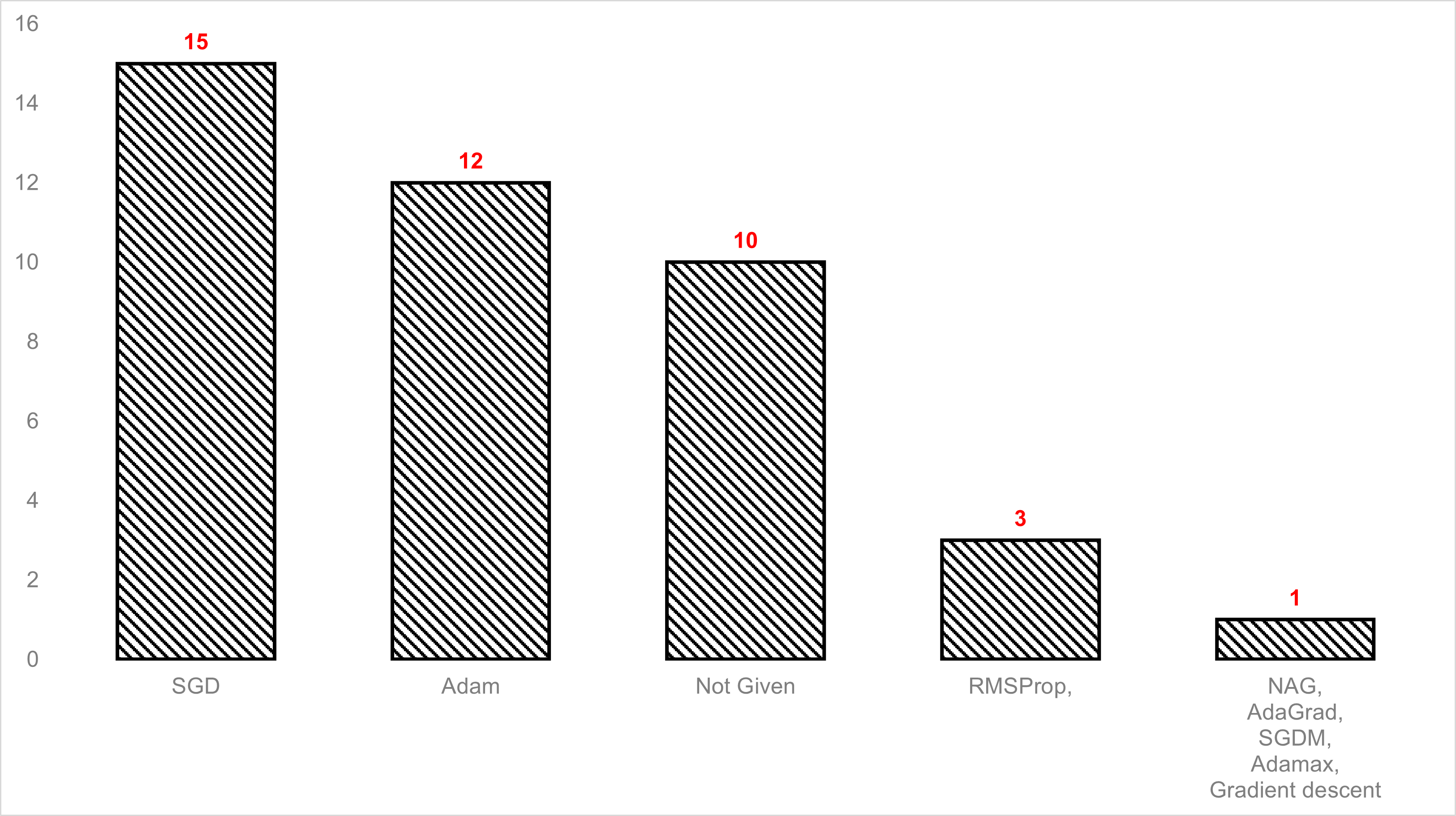}
  \centering
  \caption{The frequencies of the used optimizers.}
  \label{figure-optimizer-study}
\end{figure}

Table~\ref{table-optimizer-study} illustrates the optimizers used in the reviewed studies.
The optimizer is a significant technique used to change the weights with the learning rate in order to minimize the loss function during the neural network training. 
As shown in figure \ref{figure-optimizer-study}, the stochastic gradient descent (SGD) is the most commonly used optimizer, followed by the Adaptive Moment Estimation (Adam). 
SGD performs a fast training process as it updates a training example on each iteration. 
Some studies like \cite{Mohammadian2017Comparative} and \cite{Roshan2020Fine} propose the Adam algorithm to avoid local minimums and to converge faster than SGD. 
Furthermore, the deficiency in some of the mentioned studies \cite{Vo2016New, Lam2018Retinal, Lian2018Deep, Hazim2018Early, Li2019Diagnostic, Math2019Identification, Yip2019Enhanced, Dekhil2019Deep, Qomariah2019Classification, Mateen2020Exudate} is that they do not specify the optimizers used during the training. 
However, few studies used different optimizer algorithms such as Root Mean Square Error Probability (RMSProp), Numerical Algorithms Group (NAG), Adaptive Gradient Algorithm (AdaGrad), Stochastic Gradient Descent with Momentum (SGDM), and Adamax.

\begin{table}[!htb]
\centering
\caption{Optimization methods used in the review studies.}
\label{table-optimizer-study}
\resizebox{\textwidth}{!}{%

\begin{tabular}{p{0.5\textwidth}p{0.5\textwidth}}

\toprule

Optimizer & Study \\
\midrule

Stochastic Gradient Descent (SGD) & 
\cite{Gulshan2016Development},
\cite{Choi2017Multi},
\cite{Mohammadian2017Comparative},
\cite{Li2017Convolutional},
\cite{Zhou2018Deep},
\cite{Chen2018Detection},
\cite{Lam2018Automated},
\cite{Wan2018Deep},
\cite{Xu2018Improved},
\cite{Wijesinghe2019Transfer}, 
\cite{Li2019Automatic},
\cite{Hagos2019Transfer},
\cite{Laxmi2020Adaptive},
\cite{Hacisoftaoglu2020Deep},
\cite{Wang2020Multi} \\

Adaptive Moment Estimation (Adam) & 
\cite{Mohammadian2017Comparative},
\cite{Gao2018Diagnosis},
\cite{Lam2018Automated},
\cite{Hathwar2019Automated},
\cite{Kassani2019Diabetic},
\cite{Wijesinghe2019Transfer}, 
\cite{Ahmad2019Deep},
\cite{Zeng2019Automated},
\cite{Bodapati2020Blended},
\cite{Roshan2020Fine},
\cite{Yu2020Performance},
\cite{Wang2020Multi} \\

Not Given & 
\cite{Vo2016New},
\cite{Math2019Identification},    
\cite{Yip2019Enhanced},    
\cite{Qomariah2019Classification},    
\cite{Mateen2020Exudate},
\cite{Lam2018Retinal},
\cite{Lian2018Deep},
\cite{Hazim2018Early},
\cite{Li2019Diagnostic},
\cite{Takahashi2017Applying} \\

Root Mean Square Error Propability (RMSProp) & 
\cite{Lam2018Automated},
\cite{Voets2019Reproduction},
\cite{Zhang2019Automated}\\

Numerical Algorithms Group (NAG) & 
\cite{Lam2018Automated} \\

Adaptive Gradient Algorithm (AdaGrad) & 
\cite{Lam2018Automated} \\

Stochastic Gradient Descent with Momentum (SGDM) & 
\cite{Wang2018Diabetic} \\

Adamax & 
\cite{Islam2019DeepDR} \\ 

Gradient descent &
\cite{Dekhil2019Deep} \\

\bottomrule
\end{tabular}%
}
\end{table}

\section{Conclusion}
\label{section-conclusion}

DR is one of the complications of diabetes that occurs when the blood vessels in the retina are damaged due to diabetes.
DR, if left untreated, could cause blindness. 
Early diagnosis of DR can effectively prevent visual impairment. 
Automatic classification of DR images can effectively assist physicians in diagnosing process and improves diagnostic efficiency accordingly. 
In recent years, DNN algorithms, mostly CNN architectures, have been frequently used in studies related to the analysis and classification of retinal images. 
However, since CNN training requires large datasets for high performance, transfer learning approaches are preferred due to insufficient sized medical data. 
Accordingly, in this review, \countOfReviewedPapers{} papers that performed transfer learning approaches for DR detection have been analyzed in terms of the used dataset, used architecture, used performance metrics, used optimizers, and applied preprocessing and augmentation methods. 
Findings and statistics have been summarized in the aforementioned sections.

The findings of this review are as follows:

\begin{itemize}
    \item The most studied dataset in the reviewed studies is Diabetic Retinopathy Detection Dataset on Kaggle. 
    \item The most preferred pre-trained CNN architecture is Inception-v3.
    \item Through all performance metrics, most performance measurements are made using accuracy and sensitivity (recall).
    \item Among the parameter optimization techniques, SGD is the most preferred one, followed by Adam.
    \item Additionally, in many of the reviewed studies, the used parameter optimization technique is not mentioned.
    \item Since pre-trained CNN architectures have specific input shapes, resize/reshape/rescale pre-processing method is the most implemented technique.
    \item Though data augmentation techniques are not applied (or given) in most studies, rotation is the most used technique among the given ones.
    
\end{itemize}

\subsection*{Acknowledgements}

This research did not receive any specific grant from funding agencies in the public, commercial, or not-for-profit sectors.













\bibliographystyle{elsarticle-num}

\end{document}